\documentclass[conference]{IEEEtran}
 \usepackage{amsmath,amssymb}
 \usepackage{subfigure}
 \usepackage{graphicx,graphics,color,psfrag}
 \usepackage{cite,balance}
 \usepackage{caption}
 \allowdisplaybreaks
 \usepackage{algorithm}
 \usepackage{accents}
 \usepackage{amsthm}
 \usepackage{bm}
 \usepackage{algorithmic}
 \usepackage[english]{babel}
 \usepackage{multirow}

 \usepackage{enumerate}
 \usepackage{cases}
 \usepackage{stfloats}
 \usepackage{dsfont}
 \usepackage{color,soul}
 \usepackage{amsfonts}
 \usepackage{cite,graphicx,amsmath,amssymb}
 \usepackage{subfigure}
 \usepackage{fancyhdr}
 \usepackage{hhline}
 \usepackage{graphicx,graphics}
 \usepackage{array,color}

 \usepackage{amsmath}


\newcounter{parentnumber}

\newtheorem{proposition}{Proposition}

\newtheorem{remark}{\bf Remark}
\def\phi{\varphi}

\def\l{\left}
\def\r{\right}
\def\({\left(}
\def\){\right)}

\def\b0{{\mathbf{0}}}

\newcommand{\nn}{\nonumber}

\begin{document}

\title{\huge {Joint Frequency-and-Phase Modulation for Backscatter-Tag Assisted Vehicular Positioning}}
\author{\vspace{-3pt}Kaifeng Han$^\dag$, Seung-Woo Ko$^\ddag$, Seungmin Lee$^\S$,  Woo-Suk Ko$^\S$, and Kaibin Huang$^\dag$\\
\vspace{-3pt} {\small $\dag$ Dept. of EEE, The  University of  Hong Kong, Hong Kong }\\{\small $\ddag$ Dept. of EEE, Korea Maritime and Ocean University, S. Korea}\\{\small $\S$ LG Electronics, S. Korea}\\ \vspace{-10pt} {\small Email: huangkb@eee.hku.hk}}
\maketitle

\begin{abstract}
\emph{Autonomous driving} (auto-driving) has been becoming a killer technology for next generation vehicles, whereas some fatal accidents grow concerns about its safety. A fundamental function for safer auto-driving is to recognize the vehicles' locations, termed \emph{vehicular positioning}. The state-of-the-art vehicular positioning is to rely on anchors that are stationary objects whose locations are known, i.e. satellites for GPS and base stations for cellular positioning. It is important for reliable positioning to install anchors densely, helping find enough anchors nearby. For the deployment to be cost-effective, there are some trials to use backscatter tags as alternative anchors
by deploying them on a road surface, but its gain is limited by several reasons such as short contact time and difficulties in maintenance. Instead, we propose a new {backscatter-tag assisted vehicular positioning system} where tags are deployed along a roadside, which enables the extension of contact duration and facilitates the  maintenance. On the other hand, there is a location mismatch between the vehicle and the tag, calling for developing a new backscatter transmission to {estimate their relative position}. To this end, we design a novel waveform called  \emph{joint frequency-and-phase modulation} (JFPM) for {backscatter-tag assisted vehicular positioning} where a transmit frequency is modulated for the distance estimation assuming that the relevant signal is clearly differentiable from the others while the phase modulation helps the differentiation. The JFPM waveform leads to exploiting the maximum \emph{Degree-of-Freedoms} (DoFs) of backscatter channel in which multiple-access and broadcasting channels coexist, leading to more accurate positioning verified by extensive simulations.
\end{abstract}

\section{Introduction}

There is no doubt that \emph{autonomous driving} (auto-driving) will be a key technology in next generation vehicles, which is expected to change our lives by
providing various benefits such as time savings and relaxation during driving and realize the vision of smart city \cite{choi2016millimeter2}.
On the other hand, there is growing concern about its safety due to a few fatal accidents e.g., Tesla's Autopilot crash and Uber's Arizona crash. To enable safe auto-driving, the most fundamental task is to know vehicles' locations termed \emph{vehicular positioning}~\cite{han2018sensing}.

The state-of-art vehicular positioning is an \emph{anchors-aided approach}. An anchor is a kind of infrastructure whose location is fixed and known i.e., satellites for \emph{Global Positioning System} (GPS) \cite{cui2003autonomous} and base stations for cellular positioning \cite{Fischer2014}. After receiving reference signals from multiple anchors, a vehicle can obtain its accurate position from the estimation of relevant information such as arrival timings and angles, provided that \emph{Line-of-Sight} (LoS) links exist between the anchors and the vehicle.
It is well-known that the positioning accuracy is improved as more anchors in LoS are detected.
In other words, finding many LoS anchors is a key to reliable positioning.
One straightforward approach is to deploy anchors more densely, but it is not practical because high cost is expected.

For the approach to be cost-effective, using backscatter tags has emerged to be a viable solution as alternative anchors instead of satellites and base stations. Backscatter radio, a.k.a. \emph{Radio-Frequency Identification} (RFID), enables a tag to deliver information by backscattering and modulating an incident \emph{radio-frequency} (RF) wave \cite{zhu2018inference}. The backscatter tag thus requires no oscillator and RF component, making it possible to manufacture them with small form factors and low costs. One example is to deploy  backscatter tags on road surface as in \cite{jing2016efficient} such that when moving on the tag, a vehicle is able to read the tag's identification which is equivalent to the corresponding position.
However, this approach has some limitations in practice as follows. First, the vehicle with high velocity frequently fails to read tags' information because of the short contact duration. Second, the tags on the the road surfaces are likely to be fragile and their maintenances are difficult since heavy vehicles frequently press the tags.

\begin{figure}[t]
\vspace{-10pt}
\centering
\includegraphics[width=6cm]{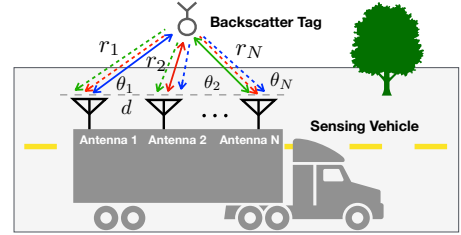}
\caption{{Backscatter-tag assisted vehicular positioning.}}\label{Fig:systemModel}
\vspace{-20pt}
\end{figure}

This paper considers a new {backscatter-tag assisted positioning system} where backscatter tags are deployed  along a roadside  (see Fig. \ref{Fig:systemModel}). Contrary to the on-road tag deployment, the distance between the tag and the vehicle becomes longer, enabling to extend the contact duration  according to the tag's cone-shape coverage \cite{Qin2017} and recent advance in long-range backscatter transmission \cite{hessar2018netscatter}. Besides, tags are expected to be more durable  because vehicles do no pass over them. On the other hand, the location of the tag does not correspond to that of vehicle, and their location difference {(relative position) is required to be estimated together along with the tag's location}. This is challenging in the backscatter system due to the tag's primitive signal processing capability and passive operation. To this end, we design a novel waveform called \emph{joint frequency-and-phase modulation} (JFPM) for the backscatter positioning to ensure the estimation of all distances between vehicle's antennas and the reader embedded in the corresponding individual signal paths.
A frequency of the incident wave is modulated for the distance estimation under the key assumption that
the relevant signal is observable without interference, which can be realized by phase modulation.
Given the JFPM, the backscatter-tag assisted channel can be fully exploited to provide the maximum \emph{Degree-of-Freedoms} (DoFs), leading to the improvement of positioning accuracy. {It is noteworthy that the minimum number of antennas is two for estimating relative angles between the antennas and the tag. Contrary to conventional signal processing approaches requiring antenna arrays for an accurate angle detection, our system is more cost-effective and practical.} Simulation results show the effectiveness of the proposed backscatter vehicular positioning deployment and the JFPM waveform.

\section{System Model}\label{sec:systemModel}
In this section, we briefly introduce our backscatter vehicular positioning system
with its signal model. Then the positioning problem is formulated.

\subsection{Overview of Backscatter Vehicular Positioning}\label{sec:backPosSystem}
We consider a backscatter vehicular positioning system comprising a backscatter tag deployed along a roadside and a reader-mounted sensing vehicle equipped with $N$ antennas with $d$  inter-antenna spacing (see Fig. \ref{Fig:systemModel}).
The backscatter tag embeds a binary sequence as its \emph{tag identification} (tag-ID) that has been associated with its absolute position. 
The reader on the vehicle easily obtains
the tag-ID via \emph{backscatter communication} based on ON/OFF keying \cite{han2017wirelessly, zhu2018inference}, where two load impedances of the tag's antenna
switch with mismatch (reflective) or match (absorptive) sates to transmit `1' or `0' bit, respectively.

Besides to the tag's absolute location, \emph{backscatter ranging} is required to obtain the vehicle's absolute position, which is the main theme of this paper. It provides the vehicle's relative position with respect to the tag by measuring the distances between antennas and the tag. 
 To this end, all $N$ antennas simultaneously transmit a JFPM waveform signal to the tag.
Due to the nature of backscatter channels, each antenna receives up to $N$ signals originated from not only itself but also the other antennas, providing in total $N^2$ DoFs, each of which embeds its propagation distance explained in the sequel.
Therefore, we aim at exploiting the maximum DoF
to help distance estimations between the $N$ antennas and the tag, which contributes to a more accurate  positioning result.

\begin{figure}[t]
\centering
\includegraphics[width=7cm]{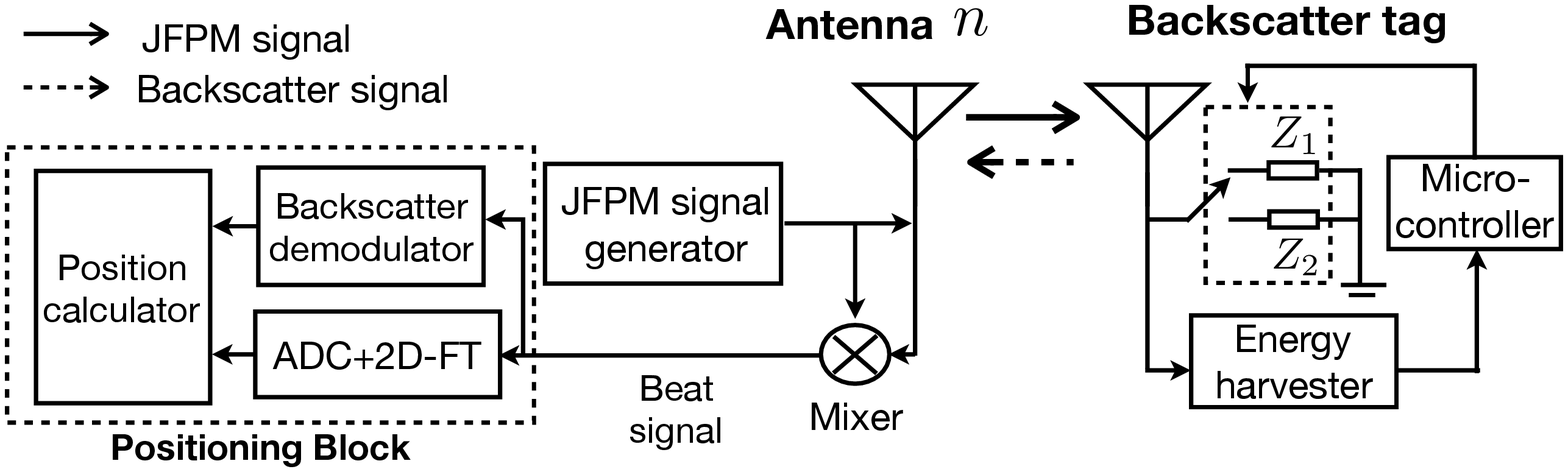}
\caption{The design of backscatter vehicular positioning system.}\label{Fig:circuitModelDesign}
\vspace{-10pt}
\end{figure}

\subsection{Signal Model}\label{sec:SignalModel}
The architecture of the backscatter vehicular positioning system is illustrated in Fig. \ref{Fig:circuitModelDesign} with the following three-step signal model.


\subsubsection{Transmit Signal at the Reader}\label{sec:transmitSignal}
Time is slotted into $L$ frames called \emph{sweeps}. The period of each sweep is $T$ and the total period is $LT$. Within the $\ell$-th sweep ($\ell \in \{1,2,\cdots, L\}$), the $n$-th antenna's JFPM waveform  is given by
\begin{align}\label{Eq:TransmitSignal}
x_n(t, \ell) = e^{j(F(t) + P_n(\ell))},\hspace{5mm} 0\leq t<T,
\end{align}
where $F(t)$ and $P_n(\ell)$ denote  frequency and phase modulation functions, respectively. It is worth highlighting that each antenna uses an equivalent frequency modulation function but a different phase modulation function due to their roles' difference. Specifically,
$F(t)$ allows the reader to estimate signals' propagation distances in a frequency domain if they are differentiable, and $P_n(\ell)$ helps the differentiation in a phase domain. The different forms of $F(t)$ and $P_n(\ell)$ will be proposed in the next section depending on the concerned driving scenarios with low or high mobilities.




\subsubsection{Backscattered Signal}\label{sec:backscatterSignal}
The received signal at the $n$-th antenna is a non-coherent combination of multiple backscatter signals originated from all $N$ antennas including both monostatic and bistatic signals, which is expressed as
\begin{align}\label{Eq:ReceiveSignal}
y_n(t, \ell) = \sum_{m = 1}^{N} A_{m,n} \cdot b \cdot x_m\l(t-\tau_{m,n},  \ell \r) + w_n,
\end{align}
where $A_{m,n}$ denotes the signal attenuation of the received signal originated from antenna $m$. The binary coefficient $b \in \{0,1\}$ refers to the transmitted bit of the tag's ON/OFF keying, $\tau_{m,n}$ denotes propagation time from $m$-th transmit antenna via the tag to the $n$-th receive antenna, and $w_n$ is the \emph{additive white Gaussian noise} (AWGN) following $\mathcal{CN}(0, \sigma^2_{\textrm{noise}})$. As shown in Fig. \ref{Fig:systemModel}, the relative distance and angle between the tag and the $n$-th antenna are denoted by $r_n$ and $\theta_n \in [0, \pi]$, respectively. The propagation time $\tau_{m,n}$ in \eqref{Eq:ReceiveSignal} is given by
\begin{align}\label{Eq:PropagationTime}
\tau_{m,n} = \frac{r_m + r_n}{c}+\frac{v(\cos\theta_n+\cos\theta_m) (t + (\ell-1)T)}{c}, 
\end{align}
where $c$ is speed of signal propagation and
the second term represents the distortion due to the vehicle's velocity $v$ resulting in Doppler shift.


\subsubsection{Beat Signal}\label{sec:beatSignal}
The tran=smit and backscattered signals are multiplied in the mixer, resulting in
\begin{align}\label{Eq:BeatSignal}
s_n (t, \ell) &= y^{\ast}_n(t, \ell) x_n(t, \ell),
\end{align}
where $(\cdot)^*$ refers to the conjugate transpose. We call it a \emph{beat signal} since
two key terms for vehicular positioning remain in the forms of vibrating frequencies defined as \emph{beat rates} (BRs). The details are given in the next section.

\subsection{Positioning Problem}
A pair of the relative distance and angle to the tag
$\{r_n, \theta_n\}$ is equivalently translated into the relative position of the $n$-th antenna, denoted by
$\textbf{z}_{n}$, which is given by
\begin{align}\label{Eq:TranslatingPosition}
\textbf{z}_{n}=-r_n (\cos \theta_n, \sin\theta_n)^\textrm{T}, \quad n=1,\cdots N.
\end{align}
The antennas' absolute positions, denoted by
$\{\textbf{p}_{n}\}_{n=1}^N$, is obtained by shifting the origin of the 2D coordinates into the tag's absolute position
$\textbf{p}_{\textrm{tag}}$ acquired by backscatter communication,  namely,
$\{\textbf{p}_{n}\}_{n=1}^N=\textbf{p}_{\textrm{tag}}+ \{\textbf{z}_{n}\}_{n=1}^N$.
We assume the backscatter communication is error-free and accurate $\textbf{p}_{\textrm{tag}}$ is always available. As a result, the vehicular positioning is reduced to the problem of discovering suitable frequency and phase  modulation functions $F(t)$ and $P_n(\ell)$  in \eqref{Eq:TransmitSignal}, enabling to distill  $\{r_n, \theta_n\}_{n=1}^N$ from $\{{s}_{n}(t,\ell)\}_{n=1}^N$.

\section{Backscatter Vehicular Positioning Technique}
This section introduces JFPM waveforms of type I and type II, which are designed for low and high mobility driving scenarios, respectively.

\subsection{Low Mobility Scenario}\label{sec:lowMobilitySection}
\begin{figure}[t]
\centering
\includegraphics[width=7.5cm]{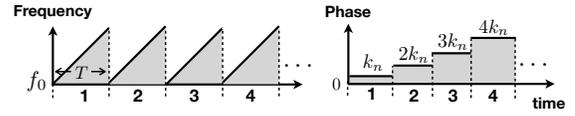}
\caption{ JFPM waveform  of type I.}\label{Fig:JFPMlowMobility}
\vspace{-10pt}
\end{figure}

In this scenario, the velocity of vehicle is very slow (e.g., parking or traffic jam) and the resultant doppler effect is marginal. We thus treat it as a stationary case (i.e., $v=0$), leading to proposing the following JFPM waveform of type I
shown in Fig.~\ref{Fig:JFPMlowMobility}.
First, the frequency modulation is given as
\begin{align}\label{Eq:FreqMod_Slow}
F(t) = 2\pi f_0 t + \pi \alpha t^2,
\end{align}
where $f_0$ is the fundamental (start) frequency and $\alpha = \frac{B}{T}$ denotes the slope that is the ratio of signal sweep bandwidth $B$ and sweep duration $T$. The resultant frequency is $f_0+\alpha t$,
which is linear increasing in each sweep.
Second, the phase modulation function is designed as
\begin{align}\label{Eq:PhaseMod_Slow}
P_n(\ell) =  2\pi k_n \ell,
\end{align}
which is constant during one sweep ($0\leq t<T$) but linearly increasing of the sweep index $\ell$.
Here, $k_n$ represents the rate of phase modulation which should be less than $1$ to avoid ambiguity. Each antenna uses different $k_n$ playing a role as a label of signals originated from antenna $n$.


Given the type I waveform, the received beat signal
is rewritten by plugging \eqref{Eq:FreqMod_Slow} and \eqref{Eq:PhaseMod_Slow} into \eqref{Eq:BeatSignal}. For the sake of brevity, we will focus on $1$-st antenna.  After ignoring the high-order term, it is approximately given as
\begin{align}\label{Eq:TwoTRX_BeatSignal_StaticCase}
s_1 (t, \ell) &\approx b \Big(\beta_{1,1} e^{j\l( 2\pi \alpha \frac{2 r_1}{c}  \r)t} \nn \\
&\hspace{0mm} + \sum_{n=2}^N\beta_{n,1} e^{j\l( 2\pi \alpha \frac{r_1 + r_n}{c}  \r)t} e^{j(2\pi(k_1 - k_n)) \ell}\Big) + \tilde{w}_1,
\end{align}
where $\beta_{1,1} = A_{1,1} e^{j\l(\frac{4  \pi f_0 r_1}{c}\r)}$ and $\beta_{n,1} = A_{n,1} e^{j\l(\frac{2 \pi f_0(r_1+r_n) }{c}\r)}$ are  constant coefficients for the monostatic signal of $x_1(t, \ell)$ and the bistatic signals of $x_n(t, \ell)$, respectively. The term $\tilde{w}_1$ represents the thermal noise after passing the mixer.

As aforementioned in Sec.~\ref{sec:beatSignal}, there are two kinds of BRs embedded in  \eqref{Eq:TwoTRX_BeatSignal_StaticCase}. The first one, determined by the propagation distances $2r_1$ and $\{r_1+r_n\}$, affects the periodicity in variable~$t$. It is defined as a \emph{BR-in-$t$} (BRiT). It is possible to extract multiple BRiTs via an \emph{one-dimensional Fourier transform} (1D-FT) over $t$. On the other hand, if there exist such antennas $n_1$ and $n_2$ satisfying $r_{n_1} = r_{n_2}$, both of the corresponding signals arrive simultaneously, making it difficult to differentiate them via the 1D-FT. To overcome the limitation, we  utilize a \emph{BR-in-$\ell$} (BRiL), defined as the periodicity in variable~$\ell$ represented by the difference of phase modulation rates  $k_1 - k_n$. It is shown that  each bistatic signal has different BRiL while monostatic signal has no BRiL. As a result, it is always available to decompose all monostatic and bistatic signals via \emph{two-dimensional FT} (2D-FT) over  $t$ and $\ell$.
The detailed procedure is  as follows.




\subsubsection{Digitalization}\label{ADCSec}
Digitalize the beat signal $s_1 (t,  \ell)$ by \emph{analog-to-digital converter} (ADC) with sampling rate $f_{\textrm{samp}}$~as
\begin{align}\label{Eq:beatSignalADC}
s_1(q, \ell) = \sum_{n=1}^{N}  \beta_{n,1} \cdot b \cdot e^{j\mathcal{F}_{n,1} q} e^{j\mathcal{P}_{n,1} \ell },
\end{align}
where $q\in\{1,\cdots, Q\}$ with $Q=T f_{\textrm{samp}}$ number of sampling points in one sweep, $\mathcal{F}_{n,1} =  2\pi \alpha \frac{r_1 + r_n}{c f_{\textrm{samp}}}$ and $\mathcal{P}_{n,1} = 2\pi(k_1 - k_n)$ are the corresponding BRiT and BRiL, respectively.

\subsubsection{2D-FT}\label{2DFTSec}
Using the digitalized beat signal \eqref{Eq:beatSignalADC}, 2D-FT is computed~as
\begin{align}\label{Eq:beatSignal2DFFT}
&\mathbf{S}_1(\mathsf{F}, \mathsf{P})=\textsf{FT}_{\textrm{2D}}\l(\{ s_1 (q,  \ell)\}, \mathsf{F}, \mathsf{P}\r)\nn \\
&= \sum_{\ell=1}^{L}  \sum_{q=1}^{Q}  \sum_{n=1}^{N} \beta_{n,1} b \cdot  e^{j\l(\mathcal{F}_{n,1} - \frac{2\pi\mathsf{F}}{Q}\r) q} e^{j \l(\mathcal{P}_{n,1} -  \frac{2\pi\mathsf{P}}{L} \r)\ell},
\end{align}
where $1\leq\mathsf{F}\leq Q$ and $1\leq\mathsf{P}\leq L$.
Note that the period of one sweep is very short ($T \leqslant 30 \mu s$ \cite{winkler2009novel}) compared with the duration of one backscatter bit \cite{van2018ambient}. Therefore, when $b=1$ (i.e., reflective state of tag), the 2D-FT is available since sufficient number of sweeps ($L \gg 1$) can be collected.

\subsubsection{Distance Estimation}\label{DistEstSec}
In the 2D-FT matrix~$\mathbf{S}_1$, up to $N$ number of peaks can be detected as $\{(\mathsf{F}_{\textrm{peak}}^{(n)}, \mathsf{P}^{(n)}_{\textrm{peak}})\}=\{( \frac{\mathcal{F}_{n,1} Q}{2\pi}, \frac{\mathcal{P}_{n,1}  L}{2\pi})\}_{n=1}^N$. Note that the reader easily matches each peak to the corresponding propagation distances by using BRiLs $\{\mathcal{P}_{n,1}\}$.
Based on the one-to-one matching, the distances  $\{r_n\}$ are estimated via the counterpart BRiTs, which are given by
\begin{align}\label{Eq:r1r2_lowMobilityCase}
\begin{cases}
r_1 = \frac{c \cdot \mathsf{F}_{\textrm{peak}}^{(1)}}{2 \alpha T},  \\
r_n =  \frac{c \cdot \mathsf{F}_{\textrm{peak}}^{(n)}}{\alpha T}-r_1, \quad n=2,\cdots, N.
\end{cases}
\end{align}
\subsubsection{Angle estimation}\label{AngEstSec}
The locations of two antennas and the backscatter tag form a triangle as shown in Fig. \ref{Fig:systemModel}.  Given the estimated $\{r_n\}$ in \eqref{Eq:r1r2_lowMobilityCase}, the counterpart angles $\{\theta_n\}$ are calculated by solving the following equation of law-of-sines on the triangle:
\begin{align}\label{Eq:LawOfSines}
\frac{(n-1)d}{\sin(\theta_1 - \theta_n)} = \frac{r_n}{\sin\theta_1} = \frac{r_1}{\sin(\pi - \theta_n)},
\end{align}
where $d$ is the spacing between adjacent antennas.

\subsubsection{Transformation to 2D-coordinates}
Given $\{r_n\}$ and $\{\theta_n\}$, the antennas' positions $\{\textbf{z}_n\}$ can be calculated by~\eqref{Eq:TranslatingPosition}.
\vskip 5pt

\begin{remark}[Effects of Antenna Numbers]\label{Remark:Gain}
\emph{Due to the nature of backscatter channels, the gains of \emph{multiple access channel} (MAC)  and \emph{broadcasting channel} (BC) coexist as follows:
\begin{itemize}
\item {\bf MAC gain}: Recall that every antenna at the reader receives not only a monostatic signal but also $(N-1)$ bistatic signals. When estimating the angle $\theta_n$, it is possible to make $\frac{\binom{N}{2}}{N} =\frac{N-1}{2}$ relevant triangle combinations. It is equivalent to $(N-1)$ equations according to \eqref{Eq:LawOfSines}, showing that more DoFs are given as $N$ increases.
\item {\bf BC gain}: A signal originated from the $n$-th antenna is broadcast to the other antennas. Following the above procedure with $N$ beat signals $\{s_n(t,\ell)\}_{n=1}^N$ makes $N$ versions of $\textbf{z}_n$, which helps mitigate estimation errors by averaging them out.
\end{itemize}}
\end{remark}

\begin{remark}[JFPM vs. FMCW]\emph{The frequency modulation \eqref{Eq:FreqMod_Slow}
has been widely used in Radar named \emph{Frequency-Modulated Continuous-Wave} (FMCW) \cite{frischen2017cooperative}. However, it cannot differentiate multiple signal paths especially when they are equally separated, resulting in the loss of DoF. The limitation can be overcome by the proposed phase modulation \eqref{Eq:PhaseMod_Slow}, leading to better positioning accuracy.   }
\end{remark}

For the type I waveform to be effective, it is essential to give a few design criterion. First, consider the maximum allowable distance to avoid ambiguity, denoted by~$r_{\max}$, which is limited by two constraints:  sampling rate $f_{\textrm{samp}}$ and the duration of one sweep $T$. Specifically, the bandwidth of beat signal depends on the maximum detectable distance, (i.e., $B_{\textrm{beat}} = \frac{\alpha 2 r_{\textrm{max}}}{c}$), which is further limited by ADC sampling rate and results in the ranging limitation: $B_{\textrm{beat}}\leq f_{\textrm{samp}} \Leftrightarrow r_{\textrm{max}} \leq \frac{f_{\textrm{samp}} c}{2 \alpha}$. Besides, since a sequence of sweeps is transmitted by each antenna, the maximum allowable delay for backscattered signals should be less than $T$ and thus $r_{\textrm{max}} \leq \frac{c T}{2}$. Combining these two provides the following proposition.
\begin{proposition}[Maximum Allowable Distance]\emph{Given bandwidth $B$, sweep duration $T$, and sampling rate $f_{\textrm{samp}}$, the maximum allowable distance between the tag and each antenna is limited by $r_{\textrm{max}}=\min \l\{\frac{f_{\textrm{samp}} c}{2 \alpha}, \frac{c T}{2}\r\}$.}
\end{proposition}

Second, it is important to determine key parameters of the phase modulation functions $\{P_n(\ell)\}$  \eqref{Eq:PhaseMod_Slow}, namely, modulation rates $\{k_n\}$ and the number of sweep $L$.
These are determined by \emph{resolution}, defined as the minimum separation between adjacent BRiLs to differentiate
multiple peaks in the same distance.
Specifically, the resolution depends on $L$ such that two BRiLs can be differentiated if the difference is larger than $\frac{2\pi}{L}$. Therefore, the following inequality should be satisfied:
\begin{align}\label{eq:ConditionSweepLowMobility}
\frac{2\pi}{L}< {\min_{n,i}\l| \mathcal{P}_{n,1} - \mathcal{P}_{i,1} \r|}\leq \frac{2\pi}{N}.
\end{align}
Here, the equality holds when adjacent BRiLs are equally separated,
 leading to the following proposition.
\begin{proposition}[Optimal Phase Modulation in the Type I Waveform] \label{Prop:LimitationOfSignalSeparation}
\emph{In the type I waveform, the optimal phase modulation function achieving the maximum resolution is
\begin{align}\label{Eq:OptimalPhaseMod}
P^*_n(\ell) = \frac{2\pi n}{N} \ell.
\end{align}
Given \eqref{Eq:OptimalPhaseMod}, a signal differentiation can be guaranteed
when the number of sweeps is larger than the number of the reader's antenna (i.e., $L>N$).
}
\end{proposition}

\subsection{High Mobility Scenario}\label{sec:highMobilitySection}
As the vehicle is moving faster, the resultant Doppler shift becomes  significant. Applying the  type I waveform thus brings about the degradation of the positioning accuracy due to the distortion specified in \eqref{Eq:PropagationTime}.  It calls for designing a JFPM waveform of type II for a high mobility~scenario.

\begin{figure}[t]
\centering
\includegraphics[width=7.5cm]{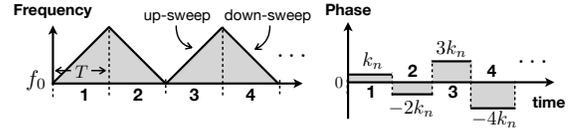}
\caption{JFPM waveform of type II.}\label{Fig:JFPMhighMobility}
\vspace{-10pt}
\end{figure}

A key idea is to use different frequency and phase modulations functions in odd and even sweeps,
making it possible to cancel out each Doppler shift assuming that it is identical to each other. Fig. \ref{Fig:JFPMhighMobility} illustrates the new JFPM waveform of type~II. Specifically, the frequency modulation adopts the triangular waveform such that odd sweeps ($\ell_{\textrm{odd}} \in \{1,3,\cdots\}$) have a \emph{linear increasing} slope, while even sweeps ($\ell_{\textrm{even}} \in \{2,4,\cdots\}$) have a \emph{linear decreasing} slope. For the phase modulation, the rate $\{k_n\}_{n=1}^N$ for odd sweeps are designed to be positive, while those of even sweeps  are negative.




Given the type II waveform, the $1$-st antenna's beat signal of an odd sweep is approximately written by ignoring high-order terms as
\begin{align}\label{Eq:BeatSignal_UpChirps}
&s_1 (t,  \ell_{\textrm{odd}}) \approx b\left[\tilde{\beta}_{1,1} e^{j\l( 2\pi \alpha \frac{2 r_1}{c} + {2\pi f_0 \frac{2 v \cos\theta_1}{c}}\r) t}e^{j\l(2\pi f_0 \frac{2 v \cos\theta_1}{c}T\r)\ell_{\textrm{odd}}}\right. \nn \\
&+ \sum_{n=2}^N\tilde{\beta}_{n,1} e^{j\l( 2\pi \alpha \frac{r_1 + r_n}{c} + 2\pi f_0\frac{v(\cos\theta_1 + \cos\theta_n)}{c}\r)t} \nn \\
&\hspace{5mm} \left. \times e^{j\l(2\pi(k_1 - k_n) + 2\pi f_0\frac{v\l(\cos\theta_1 + \cos\theta_n\r)}{c}T\r) \ell_{\textrm{odd}}}\right] + \tilde{w}_1,
\end{align}
where $\tilde{\beta}_{1,1} = \beta_{1,1}e^{-j\l(2\pi f_0  \frac{2 v \cos\theta_1}{c}T\r)}$ and $\tilde{\beta}_{n,1} = \beta_{n,1}e^{-j\l(2\pi f_0  \frac{v(\cos\theta_1 + \cos\theta_n)}{c}T\r)}$. The beat signal of an even sweep $s_1 (t,  \ell_{\textrm{even}})$ is written by replacing the positive slope $\alpha$ and phase modulation rates $\{k_n\}$ in \eqref{Eq:BeatSignal_UpChirps} with negative slope $-\alpha$ and rates $\{-k_n\}$, respectively.
It is shown that BRiT and BRiL in  both odd and even sweeps' beat signals are distorted by the same levels of the Doppler shifts, which can be eliminated if those in even and odd sweeps are separately estimated. To this end, we compute 2D-FTs~\eqref{Eq:beatSignal2DFFT} for even and odd sweeps separately, each of which the coordinates of peaks are shifted as much as the corresponding doppler shifts. By computing the difference between relevant peaks's coordinates, we can refine BRiT and BRiL with no doppler shifts, helping estimate accurate relative distances and angles. The other steps excluding the above is the same as those for a low mobility scenario.

\begin{remark}[Cost of Mitigating the Doppler Shift]\label{remark:CostDoppler}
\emph{
In the type II waveform, only a half points are given for 2D-FT than the type~I due to the separated computations in odd and even sweeps. It results in the half resolution, which is the cost paid for mitigating the Doppler shift such that the inequality \eqref{eq:ConditionSweepLowMobility} becomes
$\frac{2\pi}{\frac{L}{2}}<\frac{2\pi}{N}$, As a result, the design criterion of the number of sweeps $L$ is revised as follows.}
\end{remark}
\begin{proposition}[Number of Necessary Sweeps in the Type II Waveform] \label{Prop:LimitationOfSignalSeparationHighMob}
\emph{In the type II waveform, the signal differentiation can be guaranteed when the number of sweeps should be strictly larger than the twice number of antenna  (i.e., $L>2N$).
}
\end{proposition}

\section{Simulation Results}\label{sec:simulation}
The simulation parameters are set as follows unless specified otherwise. We set $f_0 = 24$GHz, $B = 200$MHz, $T = 30$$\mu$s, $f_{\textrm{samp}} = 5$MHz, and $A_{m,n} = 10^{-1}$. The transmit \emph{signal-to-ratio} (SNR) and the number of sweeps $L$ are $20$ (dB) and $80$, respectively.
The performance metric is a positioning error defined as the average Euclidean squared distance between estimated and ground true antennas' positions. 

\begin{figure}[t]
\centering
\includegraphics[width=6cm]{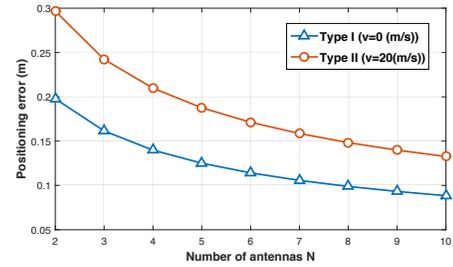}
\caption{Effect of number of antennas on positioning error.}\label{Fig:Error}
\vspace{-10pt}
\end{figure}

Fig. \ref{Fig:Error} represents the effect of number of the reader's antennas $N$ on positioning error for the JFPM waveforms of type I and type II.
The vehicle's velocity $v$ is set to $0$ and $20$ ({m}/{s}) for the type I and type II, respectively.
 It is shown that the positioning error is reduced when more antennas are deployed due to MAC and BC gains  in Remark~\ref{Remark:Gain}
and finally almost reaches $0.1$ (m), which is 3GPP's vehicular positioning requirement stated in \cite{PositioningRequirement}. In addition, the error difference between the two is around $0.05$ (m) regardless of the antenna number, referring to the resolution loss paying for the cost of eliminating Doppler effect as discussed in Remark \ref{remark:CostDoppler}.



\begin{figure}[t]
\centering
\includegraphics[width=6cm]{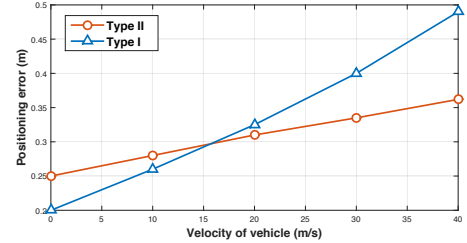}
\caption{Effect of velocity on positioning error. Two antennas are deployed at reader ($N=2$).}\label{Fig:Velocity}
\vspace{-10pt}
\end{figure}

Fig. \ref{Fig:Velocity} represents the positioning errors of the type I and type II waveforms under different velocity settings ranging from $0$ to $40$ (m/s) when two antennas are deployed at reader ($N=2$). In case of low velocity, the type I outperforms the type II  because the Doppler shift is negligible and its cancelation is unnecessary. As the velocity becomes higher, on the other hand, the Doppler shift becomes critical, yielding that the performance degradation of the type II is less than the that of the type I with the crossing point of $v=15$ (m/s).
It provides the selection between the two depending on velocity information, which is easily obtained from other existing methods.

\section{Concluding Remarks}\label{sec:conclusionRemarks}
{A novel and efficient backscatter-tag assisted vehicular positioning technique} is proposed for estimating the absolute position of vehicle. We design two novel JFPM waveforms to cope with both low and high mobility driving scenarios. Presently, we are extending the technique to mitigate severe interference signals due to the reflections of scatterers around tag and implementing the proposed system. Considering the multi-vehicle positioning scenario to investigate scheduling issue is also one promising direction.

 \vskip -30pt
\section*{Acknowledgment}
This work was supported by LG Electronics.
 \vskip -30pt

\bibliographystyle{ieeetr}

\end{document}